\documentclass[preprint,12pt,longnamesfirst,epsf]{aastex}

\begin{document}

\title{Hard X-ray Emission Associated with White Dwarfs II.}

\author{You-Hua Chu, Mart\'{\i}n A.\ Guerrero, Robert A.\ Gruendl,
Ronald F. Webbink}
\affil{Department of Astronomy, University of Illinois,
    1002 W. Green Street,
    Urbana, IL 61801; chu@astro.uiuc.edu, mar@astro.uiuc.edu, 
    gruendl@astro.uiuc.edu, webbink@astro.uiuc.edu}


\begin{abstract}

We have previously conducted a search for X-ray sources coincident 
with white dwarfs using the white dwarf catalog compiled by 
\citet{MCC99} and the {\it ROSAT} sources in the WGACAT (Paper I).
To include the white dwarfs discovered since 1999 and to include
the X-ray sources detected in {\it ROSAT} Position Sensitive 
Proportional Counter (PSPC) observations made
with a boron filter, we have carried out another search using an 
updated list of white dwarfs and the final catalogs of the
{\it ROSAT} PSPC observations with and without a boron filter.
Forty-seven new X-ray sources convincingly coincident with white 
dwarfs are found and reported in this paper.
Among these, only 5 show hard X-ray emission: three possess confirmed
or suggested late-type companions, one is apparently single, and the 
other is likely a misclassified BL Lac object.
The apparently single white dwarf with hard X-ray emission,
KPD\,0005+5106, was discussed extensively in Paper I.
Photospheric origin for the hard X-ray emission from hot DO and DQZO
white dwarfs remains a tantalizing possibility, but high-quality near
IR spectroscopic observations and monitoring of the H$\alpha$ emission
line are needed to rule out the existence of a faint dMe companion.

\end{abstract}


\newpage
\section{Introduction}

White dwarfs are not expected to emit X-rays at energies greater 
than 0.5 keV, but numerous white dwarfs appear to be associated 
with hard X-ray emission.
Most of these white dwarfs with hard X-ray emission are in binary 
systems, and the hard X-rays originate from either the accretion 
of material from the companion onto the white dwarf or simply from 
the coronal activity of a late-type companion.
A small number of white dwarfs with hard X-ray emission
appear to be single, however.
The most notable examples are WD\,0005+511 (= KPD\,0005+5106)
and WD\,1159$-$034 (= PG\,1159).
Their spectral types, DO and DQZO.4, suggest that the hard
X-ray emission may be the high-energy Wien tail of emission from
deep in the stellar atmosphere \citep[][hereafter Paper I]{Oetal03}.

In Paper I, we made a systematic search for white dwarfs with
hard X-ray emission using the WGA catalog of {\it ROSAT} X-ray 
point sources \citep[hereafter WGACAT]{WGA00}\footnote{available 
at http://wgacat.gsfc.nasa.gov/wgacat/wgacat.html.} and the list 
of white dwarfs from \citet{MCC99}.
This search was rendered incomplete because the WGACAT did not 
include {\it ROSAT} Position Sensitive Proportional Counter 
(PSPC) observations made with the boron filter.
Furthermore, 214 white dwarfs discovered since 1999 were not included
in Paper I.
To improve the completeness, we have therefore repeated the search
for white dwarfs with hard X-ray emission using an updated list
of white dwarfs and the most complete catalogs of {\it ROSAT} X-ray
sources.
The extended results are reported in this paper.

\section{Search for Hard X-ray Sources Associated with White Dwarfs}

To search for hard X-ray sources associated with white dwarfs,
we first cross-correlate the white dwarf catalog with the
{\it ROSAT} X-ray source catalogs.
The most complete list of known white dwarfs in the Galaxy is 
provided by the web version of McCook \& Sion's 
catalog\footnote{available at
http://www.astronomy.villanova.edu/WDCatalog.}, which
is continuously updated to include new white dwarfs and contains
2,449 white dwarfs at present.
For the {\it ROSAT} X-ray sources, we use the final {\it ROSAT} 
Results Archive catalogs\footnote{available at 
http://wave.xray.mpe.mpg.de/rosat/catalogue}, i.e., the Second {\it ROSAT} 
Source Catalog of Pointed Observations with the PSPC (2RXP), and 
the Second {\it ROSAT} Source Catalog of Pointed Observations with 
the PSPC with Filter (2RXF).
We adopt the same criterion for positional coincidence used in 
Paper I, i.e., $\le 1'$ offset between a white dwarf and an
X-ray source.
In addition to the coincidences reported in Table 1 of Paper I, we 
find 58 new X-ray sources apparently coincident with white dwarfs.

To confirm the coincidence between the white dwarfs and
the {\it ROSAT} X-ray sources and to examine the X-ray
spectral characteristics, we retrieved the {\it ROSAT} PSPC
event files from the High Energy Astrophysics Science 
Archive Research Center at NASA's Goddard Space Flight Center.
For each PSPC observation, we extract images in the 0.1--0.5 keV
(soft) and 0.55--2.0 keV (hard) energy bands, and compared them 
to optical images from the Digitized Sky Survey to determine (1)
whether the X-ray source is coincident with the white dwarf 
or a background object projected in its vicinity, and (2) whether 
hard X-ray emission is present and coincident with the soft X-ray 
emission.

The results of our examination of the 58 white dwarfs with
cataloged {\it ROSAT} X-ray sources within 1$'$ are presented
in Table 1.  
Columns 1--3 give the identifications, common names, and spectral 
types of the white dwarfs; columns 4--6 list the corresponding X-ray
sources in the 2RXP or 2RXF catalogs, the $ROSAT$ PSPC observations 
used for the detection, and the exposure times of the PSPC 
observations; column 7 describes the positional coincidence 
between the X-ray source and the white dwarf and the detection of
hard and soft X-ray emission; and columns 8--10 present 
the counts detected in the soft (0.11--0.41 keV), medium 
(0.52--0.9 keV), and hard (0.91--2.01 keV) bands, as reported in 
the internal files of the 2RXP or 2RXF catalogs.

Our examination shows that 11 of the 58 coincidences initially
identified by cross-correlating the cataloged positions of
white dwarfs and X-ray sources are spurious.
These objects are noted in column 7 as U1--4.
``U1" is noted when the X-ray source is not convincingly
centered on the white dwarf: either the white dwarf is outside 
the point spread function of the X-ray source (WD\,0757$-$606.2,
WD\,0949+495, WD\,1103+384, and WD\,1626+409) or the X-ray 
source is centered closer to a bright star whose spectral type 
is commonly associated with coronal activity (WD\,1407$-$475).
``U2" is noted if the X-ray source is just a weak peak on a 
bright background, and thus the source identification is
ambiguous.
``U3" is noted if a large number of candidates for optical 
counterpart are present within the PSPC's point spread function.
``U4" is noted if the X-ray source is too faint to be credible.

In column 7 of Table 1, we have further noted the detection
of soft and hard X-ray emission by ``S" and ``H," based on our 
examination of the PSPC image in the 0.1--0.5 keV and 0.5--2.4 keV 
bands, respectively.
These notes should be more accurate assessments of the detection
than the automated measurements given in the 2RXP and 2RXF catalogs
especially for objects with background sources projected in the vicinity.

It can be seen from Table 1 that only five white dwarfs are 
coincident with X-ray sources (i.e., the white dwarf is well within 
the PSPC point spread function) and show hard X-ray emission.
These five white dwarfs will be individually discussed in the
next section.
Note that Table 1 of this paper and the Table 1 in Paper I
together represent the most complete summary of detection of
X-ray emission from white dwarfs in {\it ROSAT} PSPC pointed 
observations.
The list of white dwarfs detected in the {\it ROSAT} All 
Sky Survey has been reported by \citet{Fetal96}.

\section{Description of Individual White Dwarfs with Hard X-rays}

{\bf WD\,0005+511 (KPD\,0005+5106)} was not identified in the 
survey of Paper I because its PSPC observation was made with a 
boron filter, but it was discussed in detail in Paper I.  
Briefly, the DOQZ.4 white dwarf WD\,0005+511 is apparently single, 
with no near infrared excess emission, but it shows hard X-ray 
emission peaking at 1 keV. 

{\bf WD\,0220+222 (EGGR\,18)} is a visual companion of the G0 
star HD\,14784.  The X-ray spectrum is similar to those commonly
seen from stellar coronal sources.  The hard X-ray emission
is most likely associated with the G0 star HD\,14784.


{\bf WD\,0512+326 (14\,Aur\,C; KW\,Aur\,C; HD\,33959C)} has a 
dominant soft X-ray component, but a hard X-ray component peaking
at 0.8 keV is unambiguously detected.  
This X-ray source is coincident with the single-lined spectroscopic
binary 14\,Aur\,C \citep{Hetal93}, which is itself the fainter 
companion to another single-lined spectroscopic binary, 
14\,Aur\,A$=$KW\,Aur, a $\delta$\,Sct and ellipsoidal variable.
The DA1.8 white dwarf responsible for the UV and soft X-ray emission
is not detected at optical wavelengths, but is resolved at 2\farcs0
from the primary of 14\,Aur\,C in an {\it HST} WFPC2 $F185W$ UV image
\citep{Betal01}.
The primary component of 14\,Aur\,C, an F4\,V star, is in a 
short-period orbit \citep[$P=2\fd993$:][]{Wetal92,Eetal96,Vetal98} 
with an unseen companion. 
That companion is deduced (from the spectroscopic mass function and
projected rotation velocity of the F4\,V star) to be an $\sim$M0\,V
dwarf \citep{T97,Wetal01}, making it the likely source of coronal
hard X-ray emission.

{\bf WD\,0907+336} is detected by a pointed PSPC observation 
at a 48$'$ off-axis position with a very extended point spread 
function. 
A better position can be found in the {\it ROSAT} All Sky
Survey Bright Source Catalog \citep{Vetal99}, where this
X-ray source is designated as 1RXS\,J091037.2+332920.
This X-ray source has been associated with a radio source
and classified as a BL Lac object \citep{Wetal00}.
Therefore, the X-ray source is likely unrelated to a 
white dwarf.
Note that the classification of WD\,0907+336 as a DC white
dwarf based on its ``featureless" optical spectrum \citep{WB93}
was not confirmed by \citet{EW95}.
It is most likely that ``WD\,0907+336" is a misclassified 
BL Lac object.
Such misclassification is not without precedent, as 
\citet{Fetal93} have pointed out three other BL Lac 
objects that were formerly misclassified as DC white dwarfs.

{\bf WD\,2226$-$210} is the central star of the Helix Nebula.
It was not identified in the survey of Paper I because the
white dwarf catalog of \citet{MCC99} had a sign error in 
the declination, which has been corrected in the web version of
the catalog.  
WD\,2226$-$210 shows hard X-ray emission but not near infrared
excess.
Based on the temporal variations of its hard X-ray
emission and H$\alpha$ emission line profile, it has been suggested
that WD\,2226$-$210 possesses a late-type dMe companion, and that 
the dMe star's coronal activity is responsible for the hard X-ray emission 
\citep{GRU01,GUE01}.

\section{Discussion and Conclusions}

We have completed a search for {\it ROSAT} PSPC X-ray sources 
coincident with white dwarfs.
Paper I reported the results from cross-correlating the white dwarfs 
cataloged by \citet{MCC99} and the X-ray sources detected in pointed
{\it PSPC} observation without a boron filter (WGACAT).
The current paper reports PSPC X-ray sources coincident with
white dwarfs that were discovered after 1999 or white dwarfs that
are detected only in PSPC observations made with the boron filter.
Therefore, Table 1 of Paper I and Table 1 of this paper represent 
a complete summary of pointed {\it ROSAT} PSPC observations of white 
dwarfs with and without filter.

Table 1 lists 47 cases in which a PSPC source is convincingly
coincident with a white dwarf. 
Among these X-ray sources, 5 show hard X-ray emission.
Three of these have confirmed or suggested late-type companions 
that are likely responsible for the hard X-ray emission, one is
apparently single, and the other is a misclassified BL Lac object.
The fraction of white dwarfs with apparent hard X-ray emission 
in this sample (excluding the misclassified BL Lac object), 
4/46 $\sim$ 9\%, is lower than that for the sample in Paper I,
17/76 $\sim$ 22\%.  
These different detection rates are probably caused by the different
exposure times of PSPC observations.
Column 6 of Table 1 shows that most of the PSPC observations have
short exposure times, less than 10 ks.
As shown in Paper I, hard X-ray sources associated with white dwarfs 
are much finater than the soft X-ray emission from the white dwarfs
themselves; in the most interesting cases, the hard X-ray count rate
is $\ll$1\% of the soft X-ray count rate, for example,
PG\,1159.
Deep exposures are needed to detect the faint hard X-ray emission 
from white dwarfs, especially teh apparently single ones.

No apparently single white dwarfs with hard X-ray emission are 
discovered beyond those included in the discussion of Paper I.
The most intriguing conclusion of paper I is the possibility of a
photospheric origin for the hard X-ray emission from hot DO and
DQZO white dwarfs, i.e., KPD0005+5106 and PG\,1159.
To confirm this possibility, it is necessary to rule out the 
existence of a faint, late-type dMe companion.
It is possible to diagnose the existence of a dMe companion using 
the temporal and spectral properties of the hard X-ray emission
as well as the temporal variations of the H$\alpha$ emission line.
High-quality near infrared spectra may also reveal absorption lines
from a late-type companion.
These follow-up observations are essential to confirming or rejecting
the photospheric origin of hard X-ray emission from apparently
single hot white dwarfs.

\acknowledgments

We gratefully acknowledge the support of the NASA ADP grant
NAG 5-13076.
We also thank the anonymous referee for useful suggestions 
to improve the paper.
This research has made use of the SIMBAD database, operated at 
CDS, Strasbourg, France, and the Digital Sky Survey produced at 
the Space Telescope Science Institute under U.S.\ Government grant 
NAG W-2166. We have also used data product from the 2MASS, 
which is a joint project of the University of Massachusetts and 
the Infrared Processing and Analysis Center/California 
Institute of Technology, funded by NASA and NSF.

\begin{deluxetable}{lllccrcccc}
\tablenum{1}
\tablewidth{55pc}
\rotate
\tabletypesize{\scriptsize} 
\tablecaption{White Dwarfs Coincident with ROSAT PSPC X-Ray Sources}
\tablehead{
\multicolumn{1}{c}{} & 
\multicolumn{1}{c}{} & 
\multicolumn{1}{c}{} & 
\multicolumn{1}{c}{ROSAT PSPC} & 
\multicolumn{1}{c}{ROSAT} & 
\multicolumn{1}{c}{Exp.} & 
\multicolumn{1}{c}{Pos.\ Coin.} & 
\multicolumn{3}{c}{X-Ray Counts Reported in 2RXP/2RXF (Counts)} \\
\multicolumn{1}{c}{WD Number} & 
\multicolumn{1}{c}{WD Type} & 
\multicolumn{1}{c}{Common Name} & 
\multicolumn{1}{c}{Source Number} & 
\multicolumn{1}{c}{Obs. No.} & 
\multicolumn{1}{c}{(ks)} & 
\multicolumn{1}{c}{Comments\,\tablenotemark{a}} &
\multicolumn{1}{c}{0.10-0.40 keV} & 
\multicolumn{1}{c}{0.41-0.90 keV} &
\multicolumn{1}{c}{0.91-2.00 keV} \\
\multicolumn{1}{c}{(1)} & 
\multicolumn{1}{c}{(2)} & 
\multicolumn{1}{c}{(3)} & 
\multicolumn{1}{c}{(4)} & 
\multicolumn{1}{c}{(5)} & 
\multicolumn{1}{c}{(6)} & 
\multicolumn{1}{c}{(7)} & 
\multicolumn{1}{c}{(8)} &
\multicolumn{1}{c}{(9)} &
\multicolumn{1}{c}{(10)}
}
\startdata  
0005+511   & DOQZ.4  & KPD\,0005+5106    & 2RXF\,J000817.7$+$512315 & 200428 &  4.9~~~ & Good,S,H & ~172 $\pm$ 13 & ~~7 $\pm$ 3~ & ~18 $\pm$ 4~ \\
0044$-$121 & PG1159  & in NGC\,246       & 2RXF\,J004703.4$-$115214 & 200842 &  7.3~~~ & Good,S,- &~586 $\pm$ 25 & ~-2 $\pm$ 4~ & ~~0 $\pm$ 0~ \\
0055$-$279.2 & DA    & SGP\,4:46         & 2RXP\,J005724.6$-$274323 & 701223 & 46.2~~~ & U4,-,-   & ~~-4 $\pm$ 6~ & ~~3 $\pm$ 2~ & ~10 $\pm$ 3~ \\
0147+674   & DA1.6   & GD\,421           & 2RXF\,J015111.7$+$673930 & 200828 &  2.3~~~ & Good,S,- &~~34 $\pm$ 6~ & ~~0 $\pm$ 0~ & ~~1 $\pm$ 3~ \\
0203$-$138 & DA1     & GD\,1104          & 2RXP\,J020548.8$-$133817 & 700550 &  2.4~~~ & Good,S,- &~429 $\pm$ 24 & ~~0 $\pm$ 47 & ~-6 $\pm$ 33 \\
0220+222   & DA5     & EGGR\,18          & 2RXP\,J022334.0$+$222726 & 190230 &  0.9~~~ & Good,S,H &~199 $\pm$ 15 & 100 $\pm$ 10 & ~81 $\pm$ 9~ \\
0340$-$453 & DA      & QSF\,1:14         & 2RXP\,J034144.5$-$451341 & 900495 & 24.2~~~ & U4,-,-   &~~71 $\pm$ 15 & ~13 $\pm$ 5~ & ~39 $\pm$ 9~ \\
0346$-$011 & DA1.2   & GD\,50,EGGR\,288  & 2RXF\,J034850.1$-$005831 & 200425 &  1.9~~~ & Good,S,- &~537 $\pm$ 23 & ~~0 $\pm$ 0~ & ~~0 $\pm$ 0~ \\
0354$-$368 & DA1     & ...               & 2RXP\,J035630.4$-$364119 & 190502 &  2.0~~~ & Good,S,- &1370 $\pm$ 39 & ~-3 $\pm$ 2~ & ~-1 $\pm$ 3~ \\
0419$-$487 & DA8     & LTT\,1951         & 2RXP\,J042104.8$-$483903 & 190090 &  0.2~~~ & U4,-,-   &~~46 $\pm$ 7~ & ~15 $\pm$ 4~ & ~26 $\pm$ 5~ \\
0455$-$282 & DA.88   & RE\,J0457$-$280   & 2RXP\,J045712.5$-$280757 & 200484 &  0.7~~~ & Good,S,- &~~40 $\pm$ 7~ & ~~0 $\pm$ 2~ & ~~0 $\pm$ 0~ \\
0503+019   & DA.88   & HS\,0503+0154     & 2RXP\,J050540.0$+$015823 & 700826 & 10.0~~~ & Good,S,- &~856 $\pm$ 36 & ~33 $\pm$ 11 & ~~1 $\pm$ 10 \\
0512+326   & DA1.8 & 14\,Aur\,C, HD\,33959C & 2RXF\,J051523.5$+$324109 & 200815 &  2.3~~~ & Good,S,H &1130 $\pm$ 34 & ~52 $\pm$ 7~ & ~44 $\pm$ 7~ \\
0510$-$418 & DA.96   & RE\,J0512$-$414   & 2RXF\,J051223.4$-$414534 & 200805 &  6.7~~~ & Good,S,- &~590 $\pm$ 25 & ~-2 $\pm$ 4~ & ~~0 $\pm$ 0~ \\
0549+158   & DA1.5   & GD\,71, EGGR\,210 & 2RXF\,J055227.7$+$155320 & 200423 &  3.0~~~ & Good,S,- &~679 $\pm$ 26 & ~~0 $\pm$ 3~ & ~~0 $\pm$ 0~ \\
0631+107   & DA1.8   & RE\,J0633+104     & 2RXP\,J063349.1$+$104143 & 900355 &  2.9~~~ & Good,S,- &~524 $\pm$ 28 & ~-1 $\pm$ 23 & ~-1 $\pm$ 24 \\
0642$-$166 & DA2     & Sirius B          & 2RXF\,J064508.6$-$164240 & 200422 &  3.2~~~ & Good,S,- &3610 $\pm$ 60 & ~~0 $\pm$ 0~ & ~~0 $\pm$ 2~ \\
0757$-$606.2 & DA1   & in NGC\,2516      & 2RXP\,J075756.5$-$605029 & 201754 & 32.2~~~ & U1,-,H   &~~~4 $\pm$ 5~ & ~24 $\pm$ 5~ & ~13 $\pm$ 4~ \\
0814+569   & DA1.5   & PG\,0814+569      & 2RXP\,J081826.8$+$564504 & 200970 &  7.1~~~ & Good,S,- &~116 $\pm$ 12 & ~-2 $\pm$ 1~ & ~-3 $\pm$ 1~ \\
0907+336   & DC?     & ...               & 2RXP\,J091040.1$+$332931 & 900327 & 12.1~~~ & Good,S,H &1489 $\pm$ 56 & 273 $\pm$ 22 & 257 $\pm$ 21 \\
0912+536   & DXP7    & GJ\,339.1, EGGR\,250& 2RXP\,J091605.3$+$532626 & 200561 &  8.1~~~ & U4,-,-   &~~~2 $\pm$ 4~ & ~~5 $\pm$ 3~ & ~~7 $\pm$ 3~ \\
0949+495   & DA3.4   & HS\,0949+4935     & 2RXP\,J095243.0$+$492015 & 700046 &  2.9~~~ & U1,-,-   &~~64 $\pm$ 9~ & ~~7 $\pm$ 3~ & ~~3 $\pm$ 2~ \\
1013$-$050 & DA.92   & RE\,J1016$-$052   & 2RXF\,J101628.7$-$052033 & 200820 &  4.2~~~ & Good,S,- &1470 $\pm$ 39 & ~~0 $\pm$ 0~ & ~~0 $\pm$ 0~ \\
1026+453   & DA1.4   & RE\,J1029+450     & 2RXF\,J102945.1$+$450709 & 200827 &  1.4~~~ & Good,S,- &~~30 $\pm$ 6~ & ~~1 $\pm$ 3~ & ~~2 $\pm$ 3~ \\
1033+464   & DA1.7   & GD\,123           & 2RXF\,J103624.6$+$460820 & 200819 &  4.8~~~ & Good,S,- &~~92 $\pm$ 10 & -24 $\pm$ 1~ & ~~0 $\pm$ 3~ \\
1034+001   & DOZ.5   & PG\,1034+001      & 2RXP\,J103703.8$-$000822 & 200744 & 10.6~~~ & Good,S,- &~~48 $\pm$ 8~ & ~-2 $\pm$ 1~ & ~-4 $\pm$ 1~ \\
1038+633   & DA2     & PG\,1038+634      & 2RXP\,J104156.8$+$630712 & 600280 &  7.1~~~ & Good,S,- &~108 $\pm$ 21 & ~13 $\pm$ 7~ & ~-2 $\pm$ 6~ \\
1041+580   & DA1.6   & RE\,J1044+574     & 2RXF\,J104445.3$+$574429 & 200823 &  3.3~~~ & Good,S,- &~~59 $\pm$ 8~ & ~-1 $\pm$ 2~ & ~~0 $\pm$ 0~ \\
1040+492   & DA1.1   & RE\,J1043+490     & 2RXF\,J104311.0$+$490219 & 200801 &  2.1~~~ & Good,S,- &~218 $\pm$ 15 & ~~0 $\pm$ 0~ & ~~3 $\pm$ 4~ \\
1103+384   & DA      & PG\,1103+384      & 2RXP\,J110548.0$+$381239 & 700513 & 30.0~~~ & U1,-,-   &~~58 $\pm$ 14 & ~-3 $\pm$ 4~ & ~17 $\pm$ 6~ \\
1109$-$225 & DA1.5 & $\beta$\,Crt\,B, HD\,97277 & 2RXF\,J111139.3$-$224936 & 200693 &  2.9~~~ & Good,S,- &~117 $\pm$ 11 & ~~0 $\pm$ 0~ & ~~0 $\pm$ 0~ \\
1113+413   & DA2     & PG\,1113+413      & 2RXP\,J111550.9$+$410240 & 700855 & 12.1~~~ & Good,S,- &~233 $\pm$ 28 & ~-9 $\pm$ 49 & ~-4 $\pm$ 46 \\
1120+439   & DA2     & RE\,J1122+434     & 2RXP\,J112255.7$+$434255 & 900383 &  8.2~~~ & Good,S,- &~334 $\pm$ 29 & ~-7 $\pm$ 66 & ~-4 $\pm$ 56 \\
1123+189   & DA.89   & RE\,J1126+183     & 2RXF\,J112619.7$+$183919 & 200822 &  1.8~~~ & Good,S,- &~~82 $\pm$ 9~ & ~~0 $\pm$ 0~ & ~~0 $\pm$ 0~ \\
1223+478   & DA1.5   & PG\,1223+478      & 2RXP\,J122558.3$+$473232 & 200537 &  5.3~~~ & Good,S,- &~~59 $\pm$ 10 & ~~3 $\pm$ 2~ & ~~6 $\pm$ 3~ \\
1226+142   & DC      & LP\,495$-$122     & 2RXP\,J122853.4$+$135950 & 700346 &  6.1~~~ & U2,-,H   &~~-2 $\pm$ 4~ & ~~5 $\pm$ 3~ & ~13 $\pm$ 4~ \\
1232+238   & DA1.1   & RE\,J1235+233     & 2RXF\,J123515.7$+$233407 & 200829 &  2.4~~~ & Good,S,- &~~16 $\pm$ 4~ & ~~2 $\pm$ 3~ & ~~4 $\pm$ 4~ \\
1249+160   & DA2     & GD\,150           & 2RXP\,J125217.3$+$154436 & 800393 & 11.0~~~ & Good,S,- &~155 $\pm$ 18 & ~~5 $\pm$ 4~ & ~-1 $\pm$ 1~ \\
1333+497   & DA1.5   & PG\,1333+497      & 2RXP\,J133520.1$+$493115 & 200964 &  2,2~~~ & Good,S,- &~~48 $\pm$ 8~ & ~~0 $\pm$ 3~ & ~~0 $\pm$ 0~ \\
1335+700   & DA1.5   & PG\,1335+700      & 2RXP\,J133617.4$+$694924 & 200584 &  1.1~~~ & Good,S,- &~~79 $\pm$ 12 & ~-1 $\pm$ 2~ & ~~2 $\pm$ 2~ \\
1407$-$475 & DA3     & BPM\,38165        & 2RXP\,J141038.9$-$474552 & 190042 &  0.1~~~ & U1,S,H   &~~11 $\pm$ 4~ & ~~8 $\pm$ 3~ & ~11 $\pm$ 4~ \\
1413+231   & DA2.5   & EGGR\,326         & 2RXP\,J141547.6$+$225705 & 800401 & 10.7~~~ & U4,-,-   &~~10 $\pm$ 6~ & ~~7 $\pm$ 3~ & ~~5 $\pm$ 3~ \\
1415+132  & DA1.5 & Feige\,93, EGGR\,107 & 2RXP\,J141739.6$+$130154 & 150053 &  3.0~~~ & Good,S,- &~107 $\pm$ 13 & ~-5 $\pm$ 13 & ~-2 $\pm$ 9~ \\
1513+442   & DA1.5   & PG\,1513+442      & 2RXP\,J151447.6$+$440135 & 200965 & 10.0~~~ & Good,S,- &~~61 $\pm$ 10 & ~~0 $\pm$ 1~ & ~-2 $\pm$ 4~ \\
1520+525   & DOQZ.4  & PG\,1520+525      & 2RXF\,J152145.9$+$522155 & 200808 &  3.9~~~ & Good,S,- &~~82 $\pm$ 9~ & ~~0 $\pm$ 0~ & ~~0 $\pm$ 0~ \\
1548+405   & DA1     & PG\,1548+405      & 2RXP\,J155035.3$+$402604 & 200950 &  3.6~~~ & Good,S,- &~~81 $\pm$ 11 & ~~0 $\pm$ 1~ & ~-1 $\pm$ 3~ \\
1626+409   & DA2.5   & PG\,1626+409      & 2RXP\,J162824.6$+$404831 & 800363 &  9.1~~~ & U1,?,?   &~~42 $\pm$ 10 & ~10 $\pm$ 4~ & ~~3 $\pm$ 2~ \\
1819+580   & DA1.1   & RE\,J1820+580     & 2RXF\,J182030.2$+$580445 & 200804 &  2.0~~~ & Good,S,- &~219 $\pm$ 15 & ~~0 $\pm$ 0~ & ~~0 $\pm$ 0~ \\
1845+019   & DA1.7   & BPM\,93487        & 2RXF\,J184739.5$+$015734 & 201360 &  1.4~~~ & Good,S,- &~124 $\pm$ 11 & ~~0 $\pm$ 0~ & ~~0 $\pm$ 0~ \\
1845+683   & DA1.3   & RE\,J1845+682     & 2RXF\,J184508.0$+$682229 & 200810 &  3.5~~~ & Good,S,- &~125 $\pm$ 12 & ~~0 $\pm$ 0~ & ~~3 $\pm$ 4~ \\
2000$-$561.1 & DA    & RE\,J2004$-$560   & 2RXP\,J200417.7$-$560246 & 200581 &  5.6~~~ & Good,S,- &~263 $\pm$ 17 & ~~0 $\pm$ 0~ & ~-1 $\pm$ 1~ \\
2004$-$605 & DA1     & RE\,J2009$-$602   & 2RXF\,J200905.2$-$602539 & 201353 &  1.5~~~ & Good,S,- &1400 $\pm$ 38 & ~~2 $\pm$ 3~ & ~~2 $\pm$ 3~\\
2151$-$307 & DA1.7   & RE\,J2154$-$302   & 2RXP\,J215457.2$-$302904 & 700924 &  3.5~~~ & Good,S,- &1174 $\pm$ 36 & ~93 $\pm$ 10 & ~47 $\pm$ 8~ \\
2152$-$548 & DA1.1   & RE\,J2156$-$543   & 2RXF\,J215620.4$-$543820 & 201355 &  1.6~~~ & Good,S,- &~891 $\pm$ 30 & ~~1 $\pm$ 3~ & ~~3 $\pm$ 4~ \\
2159$-$414 & DA.89   & ...               & 2RXP\,J220229.0$-$411438 & 200491 &  9.7~~~ & Good,S,- &~186 $\pm$ 16 & -56 $\pm$ 1~ & ~~0 $\pm$ 1~ \\
2226$-$210 & DAO     & in NGC\,7293      & 2RXP\,J222938.8$-$205015 & 900187 &  4.7~~~ & Good,S,H &~232 $\pm$ 17 & ~28 $\pm$ 5~ & ~25 $\pm$ 5~ \\
2331$-$475 & DA.97   & RE\,J2334$-$471   & 2RXP\,J233401.6$-$471421 & 200486 &  0.8~~~ & Good,S,- &~~40 $\pm$ 7~ & ~~0 $\pm$ 2~ & ~~0 $\pm$ 0~ \\
2349+286   & DA1.5    & PG\,2349+286      & 2RXP\,J235157.6$+$285525 & 200962 &  5.8~~~ & Good,S,- &~~39 $\pm$ 8~ & ~-4 $\pm$ 1~ & ~-3 $\pm$ 3~ \\
\enddata 
\tablenotetext{a}{
``Good" -- convincing coincidence;
``U1" --  the white dwarf is outside the point spread function 
       of the X-ray source or the X-ray source is centered closer to 
       a bright star whose spectral type is commonly associated with 
       coronal activity;
``U2" -- the X-ray source is just a weak peak on a bright background 
      and thus the source identification is ambiguous;
``U3" -- a large number of candidates for optical counterpart are present
      within the PSPC's point spread function;
``U4" -- the X-ray source is too faint to be credible;
``S" -- detected in the 0.1--0.4 keV band;
``H" -- detected in the 0.91--2.0 keV band.}
\end{deluxetable}

\end{document}